\documentclass[journal=jpclcd,manuscript=article]{achemso}

\usepackage[version=3]{mhchem} 
\usepackage{color}
\usepackage{xspace}
\usepackage{geometry}
\usepackage{graphicx}
\usepackage{float}
\usepackage{chemfig}
\usepackage{subcaption}
\usepackage{adjustbox}
\usepackage{amsfonts}
\usepackage{etoolbox}
\usepackage{hyperref}
\usepackage{pgf}

\usepackage[authormarkuptext=name,authormarkup=none]{changes}
\definecolor{blue}{RGB}{0, 0, 255}
\definecolor{green}{RGB}{0, 150, 0}
\definechangesauthor[name={Yorick}, color={green}]{YLAS}
\definechangesauthor[name={Ivan}, color={violet}]{IT}
\definechangesauthor[name={Tetiana}, color={magenta}]{TO}
\definechangesauthor[name={Gianluca}, color={red}]{GL}

\usepackage[normalem]{ulem} 

\pgfkeys{
  /molecules/0/.initial=\ce{MPF}\xspace,
  /molecules/14/.initial=\ce{F{-}MPF}\xspace,
  /molecules/30/.initial=\ce{CF3{-}MPF}\xspace,
  /molecules/38/.initial=\ce{N{-}MPF}\xspace,
  /molecules/39/.initial=\ce{F{,}N{-}MPF}\xspace,
  /molecules/40/.initial=\ce{CF3{,}N{-}MPF}\xspace,
  /molecules/7/.initial=\ce{TIF}\xspace,
  /molecules/15/.initial=\ce{F{-}TIF}\xspace,
  /molecules/8/.initial=\ce{CF3{-}TIF}\xspace,
  /molecules/41/.initial=\ce{N{-}TIF}\xspace,
  /molecules/42/.initial=\ce{F{,}N{-}TIF}\xspace,
  /molecules/43/.initial=\ce{CF3{,}N{-}TIF}\xspace,
  /molecules/50/.initial=\ce{MMPIF}\xspace,
  /molecules/52/.initial=\ce{F{-}MMPIF}\xspace,
  /molecules/53/.initial=\ce{CF3{-}MMPIF}\xspace,
  /molecules/49/.initial=\ce{FYT}\xspace,
  /molecules/54/.initial=\ce{F{-}FYT}\xspace,
  /molecules/55/.initial=\ce{CF3{-}FYT}\xspace,
  /molecules/60/.initial=\ce{MXT}\xspace,
  /molecules/63/.initial=\ce{F{-}MXT}\xspace,
  /molecules/64/.initial=\ce{CF3{-}MXT}\xspace,
  /molecules/65/.initial=\ce{BODIPY/M}\xspace,
  /molecules/66/.initial=\ce{F{-}BODIPY/M}\xspace,
  /molecules/67/.initial=\ce{CF3{-}BODIPY/M}\xspace,
  /molecules/16/.initial=\ce{F{-}FMT}\xspace,
  /molecules/68/.initial=\ce{FMT}\xspace,
  /molecules/69/.initial=\ce{CF3{-}FMT}\xspace,
  /molecules/12/.initial=\ce{$4’$Cl-MPF}\xspace,
  /molecules/70/.initial=\ce{F{-}$4’$Cl-MPF}\xspace,
  /molecules/71/.initial=\ce{CF3{-}$4’$Cl-MPF}\xspace,
  /molecules/79/.initial=\ce{CF3{,}F{-}MPF}\xspace,
  /molecules/80/.initial=\ce{CF3{,}F{,}N{-}MPF}\xspace,
  /molecules/81/.initial=\ce{CF3{,}F{-}TIF}\xspace,
  /molecules/82/.initial=\ce{CF3{,}F{,}N{-}TIF}\xspace,
  /molecules/83/.initial=\ce{CF3{,}F{-}MMPIF}\xspace,
  /molecules/84/.initial=\ce{CF3{,}F{-}FYT}\xspace,
  /molecules/85/.initial=\ce{CF3{,}F{-}MXT}\xspace,
  /molecules/86/.initial=\ce{CF3{,}F{-}FMT}\xspace,
  /molecules/87/.initial=\ce{CF3{,}F{-}$4’$Cl-MPF}\xspace,
  /molecules/1/.initial=\ce{Ph{-}MPF}\xspace,
  /molecules/2/.initial=\ce{iPr{-}MPF}\xspace,
  /molecules/3/.initial=\ce{tBu{-}MPF}\xspace,
  /molecules/9/.initial=\ce{tBu{-}TIF}\xspace
}

\newcommand{\molecule}[1]{\pgfkeysvalueof{/molecules/#1}}

\author{Ivan Tambovtsev}
\affiliation[University of Iceland]{Science Institute and Faculty of Physical Sciences, University of Iceland, 107 Reykjav\'{\i}k, Iceland}
\email{ivt3@hi.is}
\author{Yorick L. A. Schmerwitz}
\affiliation[University of Iceland]{Science Institute and Faculty of Physical Sciences, University of Iceland, 107 Reykjav\'{\i}k, Iceland}
\alsoaffiliation[4]{Max-Planck-Institut f\"ur Kohlenforschung, 45470 M\"ulheim an der Ruhr, Germany}
\author{Gianluca Levi}
\affiliation[University of Iceland]{Science Institute and Faculty of Physical Sciences, University of Iceland, 107 Reykjav\'{\i}k, Iceland}
\author{Darina D. Darmoroz}
\affiliation[ITMO University]{Infochemistry Scientific Center, ITMO University, 9 Lomonosova street, Saint-Petersburg, 191002, Russia}
\author{Pavel V. Nesterov}
\affiliation[ITMO University]{Infochemistry Scientific Center, ITMO University, 9 Lomonosova street, Saint-Petersburg, 191002, Russia}
\author{Tetiana Orlova}
\affiliation[ITMO University]{Infochemistry Scientific Center, ITMO University, 9 Lomonosova street, Saint-Petersburg, 191002, Russia}
\author{Hannes Jónsson}
\affiliation[University of Iceland]{Science Institute and Faculty of Physical Sciences, University of Iceland, 107 Reykjav\'{\i}k, Iceland}
\email{hj@hi.is}

\title{Fine Tuning of the Rotational Speed of Light-Driven, Second-generation Molecular Motors by Fluorine Substitution}

\begin{document}


\renewcommand*\tocentryname{TOC Graphic}
\begin{tocentry}
   \includegraphics[width = \textwidth]{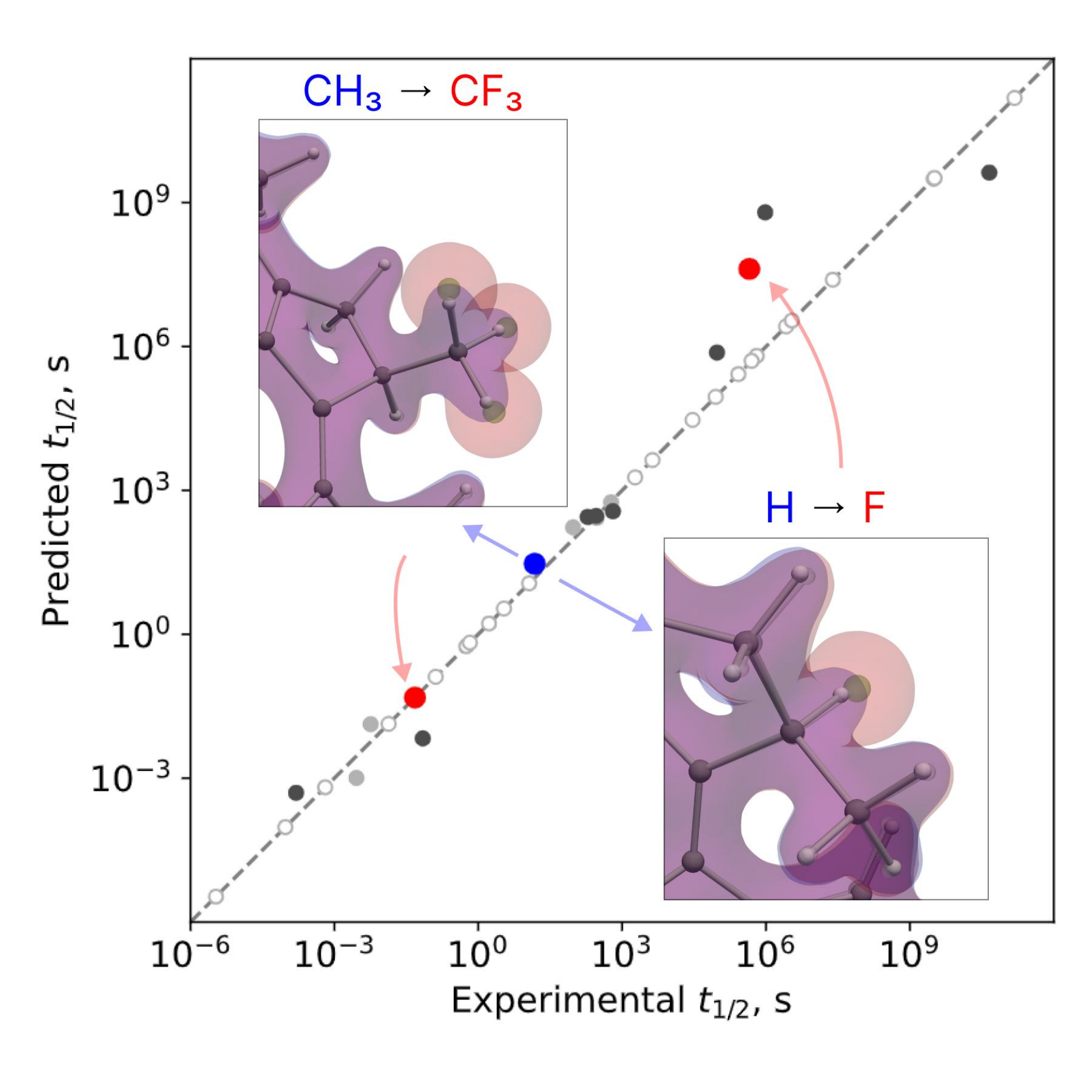}
\end{tocentry}

\begin{abstract}
The elementary steps in the rotation of several second-generation molecular motors are analysed by finding the minimum energy path between the metastable and stable states and evaluating the transition rate within harmonic transition state theory based on energetics obtained from density functional theory. Comparison with published experimental data shows remarkably good agreement and demonstrates the predictive capability of this approach. While previous measurements by Feringa and coworkers [{\it Chem.\,Eur.\,J.}\,{\bf 2017}, {\it 23}, 6643] have shown that a replacement of the hydrogen atom at the stereogenic center by a fluorine atom can slow down the rate-limiting thermal helix inversion (THI) step by raising the energy of the transition state, even to the extent that the backreaction in the ground state becomes preferred in some cases, we find that a replacement of a CH$_3$ group by CF$_3$ at the same site accelerates the THI by elevating the energy of the metastable state without affecting the transition state significantly. Since these two fluorine substitutions have an opposite effect on the rate of the THI, the combination of both can provide ways to fine tune the rotational speed of molecular motors.
\end{abstract}



\begin{figure}[H]
 \includegraphics[width = 0.65\textwidth]{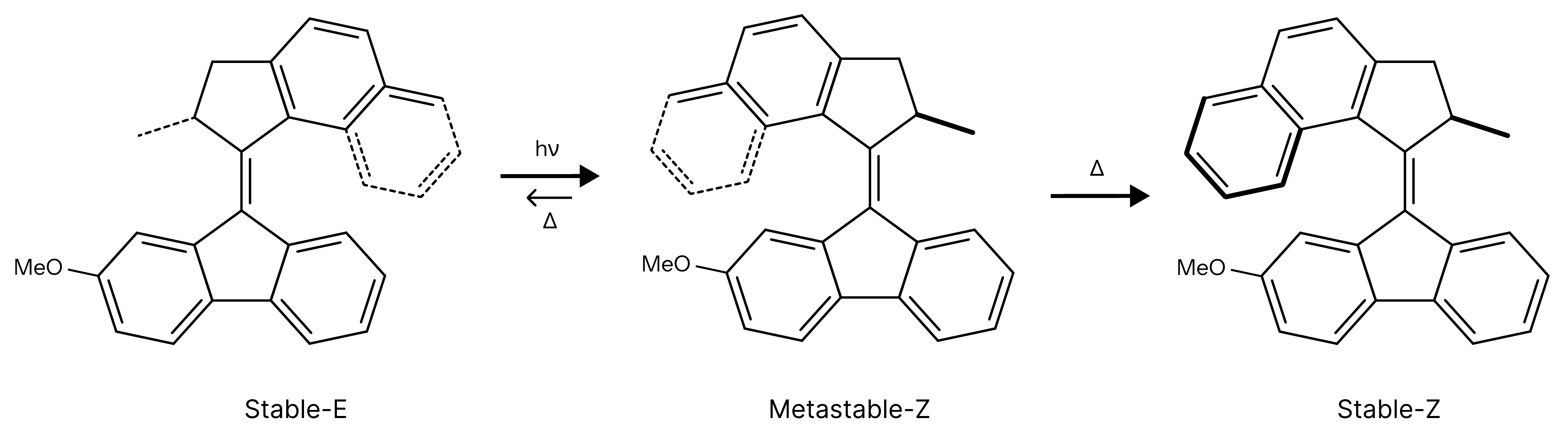}
  \caption{
  Illustration of the first two steps in the rotation of a second-generation molecular motor developed by Feringa and coworkers.\cite{vicarioControllingSpeedRotation2005}
  After photoabsorption, the structure evolves from the stable E state to the metastable Z state, first a ca.\,\,90$^{\circ}$ rotation in the excited electronic state and then further rotation after returning to the ground state. 
  From there, a thermally activated transition, the thermal helix inversion (THI), can bring the molecule to the stable Z state thereby completing a ca.\,\,180$^{\circ}$ rotation. For an efficient rotor, the transition back to the E state should have significantly lower probability than the THI. Absorption of a second photon in the stable Z state and a repeat of a THI step then completes a full circle.
  }
  \label{fig:two_ways}
\end{figure}

The development of light-powered molecular motors, where one part of a molecule rotates a full cycle with respect to the other, has been an active field of research in the past two decades, 
driven by their applications in a wide range of fields, such as optics, photonics and light-driven soft materials, to name a few. Several such molecular motors have therefore been developed.
\cite{corraPhotoactivatedArtificialMolecular2023, baronciniPhotoRedoxDrivenArtificial2020, jeongMolecularSwitchesMotors2020}. 
Most contain a C=C double bond 
with one side of the bond referred to as the `rotor' and the other referred to as the `stator'. The absorption of a photon leads to a ca.\,\,90$^{\circ}$ rotation of the double bond in the excited electronic state and, after relaxation to the ground electronic state, further rotation to a metastable state.  
From there, a thermally induced transition, the so-called thermal helix inversion (THI), to a more stable state can occur, corresponding to a rotation by 180$^{\circ}$. The stage is then set for the absorption of another photon and analogous steps as before to complete the 360$^{\circ}$ rotational cycle.
The absorbed energy from the photons is thereby converted into repeatable and controllable molecular motion.\cite{garcia-lopezLightActivatedOrganicMolecular2020,  poolerDesigningLightdrivenRotary2021, rokeMolecularRotaryMotors2018, koumuraLightdrivenMonodirectionalMolecular1999}
If, however, the rate of the THI is lower than the rate of thermally activated return to the original state on the ground state energy surface,
the molecule is not likely to perform a full rotation.
The energy landscape corresponding to the ground electronic state therefore 
plays an important role in addition to that of the electronically excited state.
An example molecule that has been shown to be an efficient rotor\cite{vicarioControllingSpeedRotation2005} is shown in figure~\ref{fig:two_ways} with an illustration of the various states, and the transitions between them.

A great deal of effort has focused on various structural modifications of the molecules
to tune both the wavelength of light capable of inducing the photochemical step of the rotation and the rate of the subsequent THI step \cite{poolerDesigningLightdrivenRotary2021, vicarioControllingSpeedRotation2005}. This has led to the development of three generations of molecular motors. 
A particularly interesting and promising one is the second generation where only a single stereogenic center in the molecular structure is sufficient to ensure unidirectional rotation. 
There, the energy barriers for the two THI steps in the rotary cycle are similar, unlike the first generation of rotors\cite{koumuraSecondGenerationLightDriven2002, conyardChemicallyOptimizingOperational2014, vicarioFineTuningRotary2006, rokeMolecularRotaryMotors2018}. 
Moreover, when the stator part of a second-generation molecular motor  
is symmetric, the third and fourth steps that take place after absorption of the second photon are essentially the same as the first and second steps. 

This makes second-generation molecular motors particularly suitable for the development of light-controllable materials where molecular-scale rotational motion is amplified to a macroscopic level due to a cooperative effect on the environment. 
Molecular motors can be used to create continuous unidirectional movement of a 
macroscopic motor, where the rotation speed is determined indirectly by 
that of the molecular motors used as dopants.\cite{orlovaRevolvingSupramolecularChiral2018}. 
Moreover, since the photoisomerization of second-generation motors is accompanied by a change in molecular chirality that is amplified to the supramolecular level, a variety of optical and photonic elements can also be developed.
\cite{yangLighttriggeredModulationSupramolecular2023,kimPhotoresponsiveChiralDopants2019}
For example,
by doping a liquid crystal (LC) or a liquid crystalline network with molecular motors,
new prospects for soft robotics and multifunctional soft actuators have been opened.\cite{houPhototriggeredComplexMotion2022,lanAmplifyingMolecularScale2022,houPhotoresponsiveHelicalMotion2021,sunLightDrivenSelfOscillatingBehavior2021}.  

The optimal rotation rate depends on the application and is different, for example, for liquid crystal mixtures, LC-based elastomers and polymers. 
Slow rotation is suitable for data storage and information technology while fast rotation is appropriate for switchable optical elements such as liquid crystal displays, sensors and optical gates. If the characteristic elastic relaxation time of a chiral nematic liquid crystal significantly exceeds the photoisomer relaxation time, then the switching time is controlled by the rotational viscosity and elastic constants of the liquid crystal host. 

The THI is typically the rate-limiting step in the rotation.
Various ways of tuning its rate have therefore been explored, mainly by modifying the structure of the molecular skeleton. One approach for tuning molecular properties that has been used in several various contexts is fluorination, i.e.\, the replacement of one or more hydrogen atoms with fluorine.
Fluorination has, for example, been used to adjust the half-life of the helical LC state of chiral azobenzenes with almost no effect on the photoabsorption properties \cite{huangLongLivedSupramolecularHelices2018,  blegerOFluoroazobenzenesReadilySynthesized2012}. 
Fluorination of second-generation molecular motors has been explored by Feringa and coworkers \cite{stackoFluorineSubstitutedMolecularMotors2017},
who found that a replacement of a hydrogen atom by a fluorine atom at the stereogenic center, i.e.\ the rotor's carbon atom next to the double bond, slows down the THI. Their analysis of experimentally measured rates and density functional theory (DFT) calculations of energy barriers led to the conclusion that the main effect of this fluorine substitution is an increase in the energy of the transition state due to steric effects while the relative free energy of the stable and metastable states is not significantly affected.\cite{stackoFluorineSubstitutedMolecularMotors2017} 
They also found that for some of the fluorinated molecules,  
the thermally activated transition from the metastable state back to the initial state becomes faster than the THI, thereby inhibiting rotation.

In the present theoretical study, another hydrogen to fluorine substitution is explored, namely the replacement of a CH$_3$ group by a CF$_3$ group at the stereogenic center. We find this to have the opposite effect, namely an increase of the THI rate and thereby increased rotational speed. The reason for this is an increase in the energy of the metastable state while the transition state turns out not to be affected significantly. Harmonic transition state theory (HTST) is used to estimate the transition rate of the thermally activated steps, with both the pre-exponential factor in the Arrhenius rate expression as well as the energy barrier obtained from density functional calculations with a hybrid density functional. 
To test this approach, we compare the calculated half-life for a wide range of previously investigated second-generation molecular motors with experimentally measured values and find remarkably close agreement. 
Calculations are also performed to assess the effect of fluorine substitution on the wavelength of light absorbed by the stable and metastable states.
In addition to fluorination, we study the effect of 
placing an N atom instead of a CH in the stator where steric hindrance occurs during the THI.



\begin{figure}[H]
    \includegraphics[width=\textwidth]{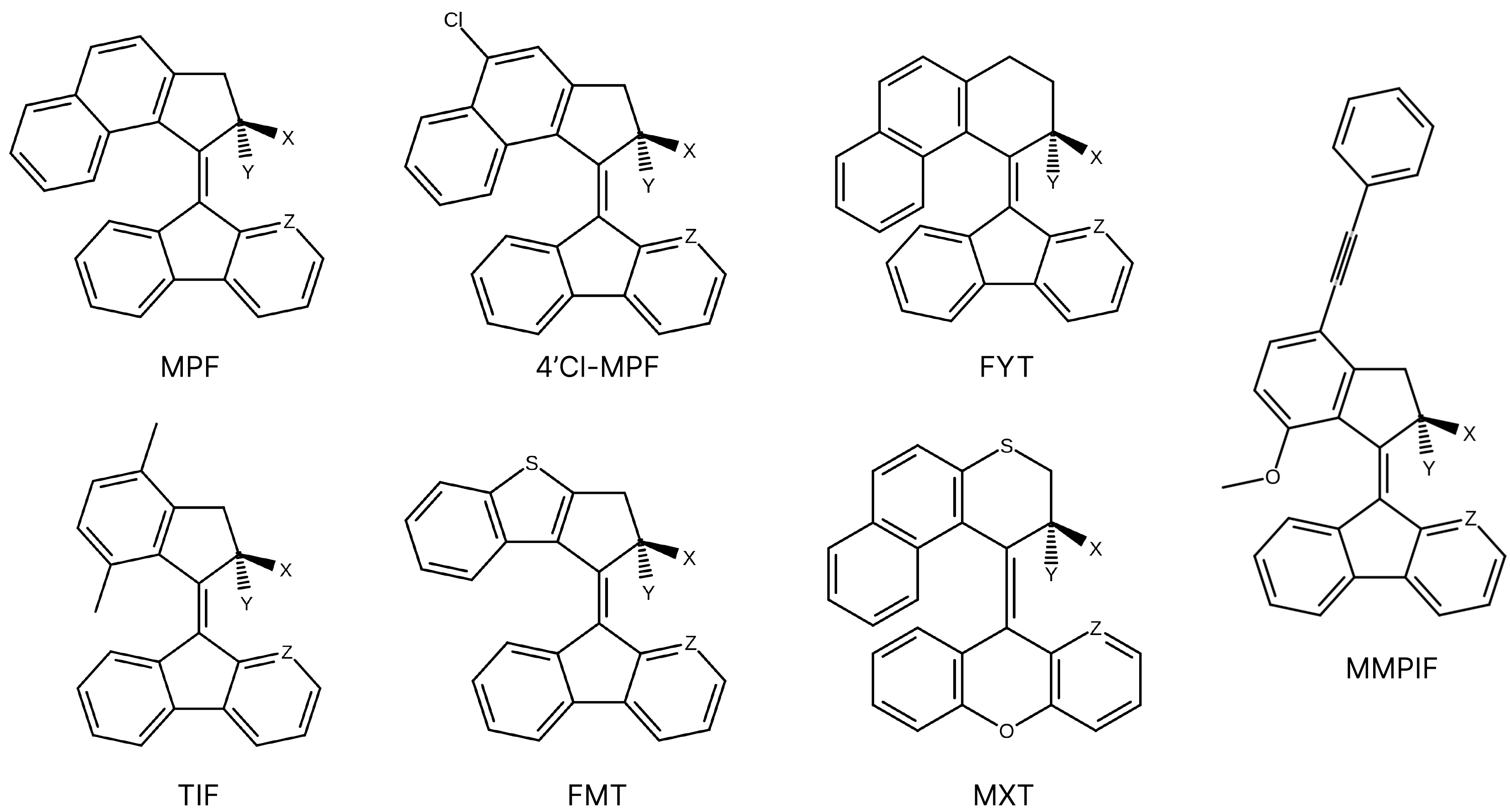}
    \caption{
    The molecular motors investigated in the present study to assess the effect of fluorine and nitrogen substitutions. The sites are labeled X, Y, and Z and they have a CH\textsubscript{3}, H, and CH, respectively, in the base molecules, but CF\textsubscript{3}, F, or N in the modified molecules. 
    }
    \label{fig:molecules}
\end{figure}

The second-generation motors studied here are shown in figure~\ref{fig:molecules}.
The sites are labeled by X, Y, Z, and have CH\textsubscript{3}, H, and CH, respectively, in the base molecule while the substitutions introduce CF\textsubscript{3}, F, or N. 
If Z is CH, the stator is symmetric and the state after THI 
is equivalent to the initial state.

\begin{figure}[H]
    \includegraphics[width = \textwidth]{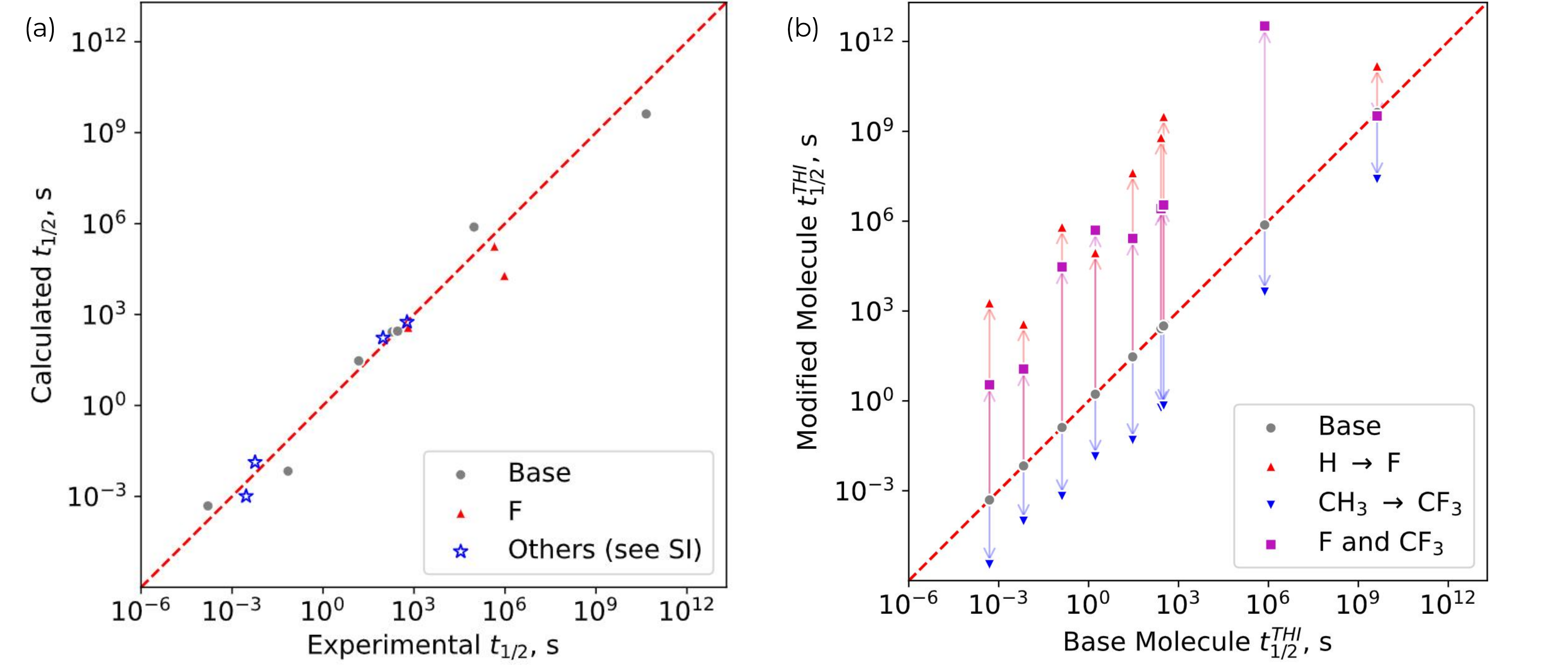}
	\caption{
  (a) Comparison of calculated and measured half-life of the metastable state of various second-generation molecular motors. 
  Both the THI step and the backreaction are taken into account.
Data corresponding to the base molecules (shown in figure 1) is shown in gray, while red illustrates data for molecules where \ce{H} has been replaced by \ce{F} at site Y. Data on additional second-generation motors is marked with blue stars.
  The experimental data is obtained from Refs.~\citenum{vicarioFineTuningRotary2006,
pollardRedesignLightdrivenRotary2008,
pollardEffectDonorAcceptor2008,
stackoFluorineSubstitutedMolecularMotors2017,
  vicarioControllingSpeedRotation2005,
  cnossenTrimerUltrafastNanomotors2009,
  feringaControlMotionMolecular2001}
  (see SI).
   The agreement between measured and calculated values is remarkably good, the red dashed line indicating perfect agreement.
 (b) Calculated half-life of the THI step and its change by
 substitution of \ce{CH3} by \ce{CF3} at site X (blue), substitution of \ce{H} by \ce{F} at site Y (red), and both substitutions combined (purple).  
 The \ce{CF3} substitution increases the rate of the THI step, whereas the \ce{F} substitution decreases it, as indicated by the arrows.
 }
	\label{fig:fit}
\end{figure}

We first address the accuracy of the theoretical approach used here for estimating the rate of the thermally activated steps, the THI and the backreaction in the ground electronic state.
Figure~\ref{fig:fit}(a) compares the calculated half-life of the 
metastable state of the various rotors with experimentally measured values.
Both the THI and backreaction steps are taken into account.
Results for 
the base molecules as well as molecules where H has been replaced by F at site Y, and several other second-generation motors for which experimental data is available are shown.  The experimental data is taken from Refs. \citenum{vicarioFineTuningRotary2006,pollardRedesignLightdrivenRotary2008,pollardEffectDonorAcceptor2008,stackoFluorineSubstitutedMolecularMotors2017,vicarioControllingSpeedRotation2005,cnossenTrimerUltrafastNanomotors2009,feringaControlMotionMolecular2001} (see SI). 
Five molecules with an alkane group at site X different from the methyl group are included in the comparison, as experimental data for them
is available.\cite{vicarioFineTuningRotary2006, bauerTuningRotationRate2014, pollardEffectDonorAcceptor2008}
The close agreement between measured and calculated values lends support for the theoretical methodology used here and thereby also the predictions we make for new modifications of the rotors.
For molecules \molecule{14} and \molecule{15}
the calculated rate of the backreaction turns out to be larger than the rate of the THI step, in agreement with experimental measurements \cite{stackoFluorineSubstitutedMolecularMotors2017}.


Figure~\ref{fig:fit}(b) shows predicted changes in the rate of the THI step
for the various molecular motors by introducing \ce{CF3} at site X and/or \ce{F} at site Y. 
In all cases, the \ce{F} substitution decreases the rate, 
consistent with the measurements for four of these molecules by Feringa and coworkers.\cite{stackoFluorineSubstitutedMolecularMotors2017} 
The introduction of \ce{CF3} is, however, predicted to increase the rate. 
The reason for the decrease in the rate by the \ce{F} substitution is higher energy barrier for the THI step, while the increase in the rate by the \ce{CF3} substitution is due to elevated energy of the metastable state.
The pre-exponential factor in the Arrhenius rate expression is typically increased by the former and reduced by the latter, but the changes are within a factor 2, so the changes in activation energy dominate.
When both substitutions are made, the half-life is in general increased, but less than for the \ce{H} to \ce{F} substitution alone. The combination of the two, therefore, enables finer tuning of the rate.
The changes in the pre-exponential factor and the activation energy are tabulated in the SI.
In most cases, the rate of the THI is greater than that of the backreaction, 
and since it is the rate-limiting step in the rotation, it defines the overall rotational speed. Only for \molecule{55} and \molecule{64} is the backreaction faster than the THI.

\begin{figure}[H]
	\includegraphics[width = 0.8\textwidth]{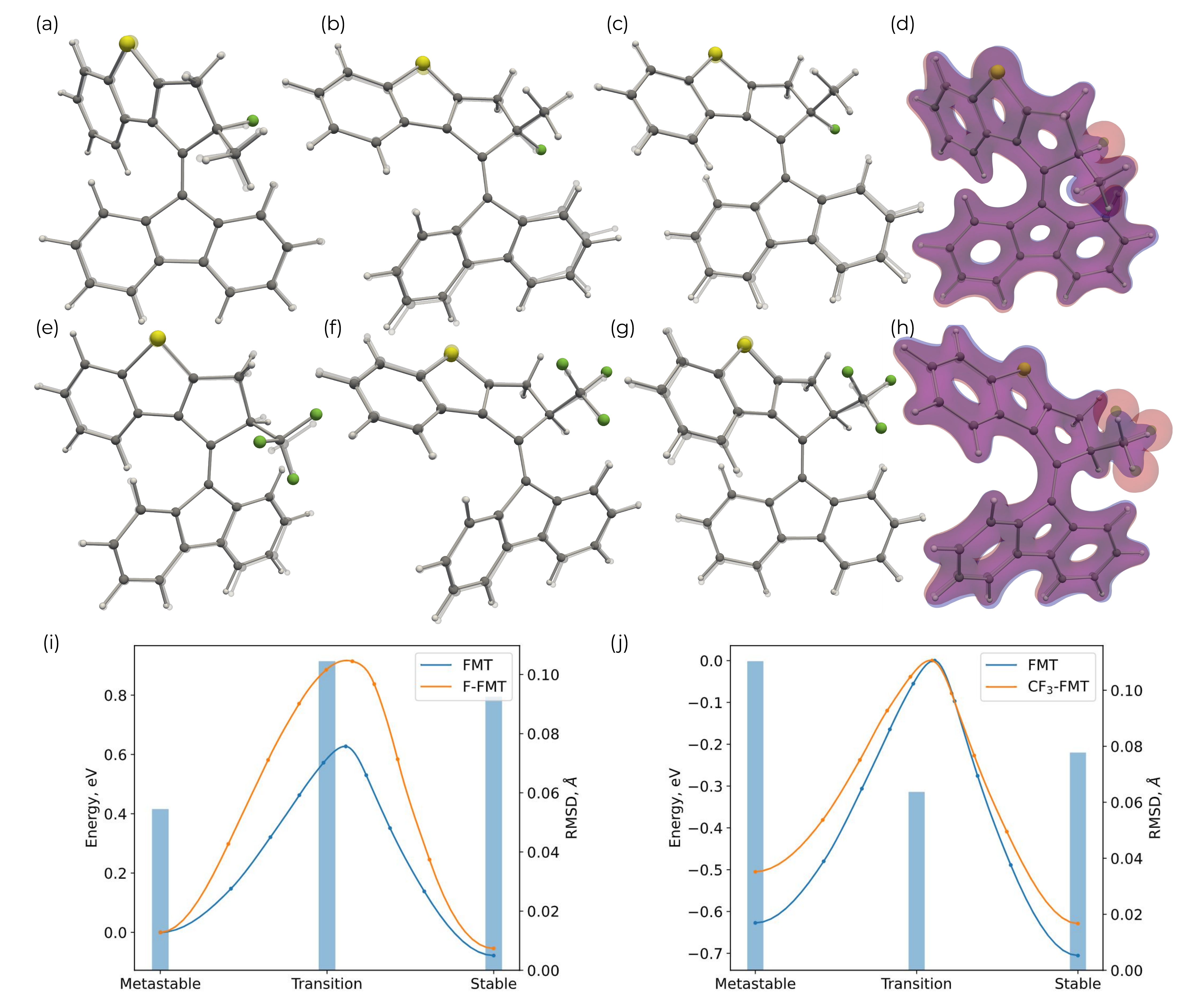}
	\caption{
    Effect of the H to F substitution at site Y, and the CH\textsubscript{3} to CF\textsubscript{3} substitution at site X of the \molecule{68} molecule.
    The F atoms are shown in green.
    (a,b,c): Change in the metastable state, transition structure, 
    and stable state due to the H to F substitution (\molecule{16}). The structures are aligned at the C=C bond.  
    (e,f,g): Analogous change due to the CH\textsubscript{3} to CF\textsubscript{3} substitution (\molecule{69}).
    The electron density rendered at 0.1 Bohr$^{-3}$ is shown for the metastable state of \molecule{16} in (d), and for the transition structure of \molecule{69} in (h). 
    Blue: original molecule, \molecule{68}. Red: Modified molecule.
    (i) and (j): Minimum energy path between the metastable and stable states for the base and modified molecules.
    The blue bars indicate the RMSD between the structure of the base and modified molecule. 
    The zero of energy is chosen to be the energy of the state with the smallest RMSD.
    The H to F substitution mainly affects the transition structure and reduces the rate of THI, but the CH\textsubscript{3} to CF\textsubscript{3} substitution mainly affects the metastable state and increases the rate.  
 }
	\label{fig:data}
\end{figure}

As can be seen in figure~\ref{fig:fit}, the rate of THI varies over a wide range, nearly 15 orders of magnitude. 
Figure~\ref{fig:data}(a-h) shows 
an analysis of the effect of the two fluorine substitutions on the atomic structure of the \molecule{68} molecule in the stable and metastable states, characterized by the two local minima, as well as in the transition state characterized by the saddle point on the energy surface, i.e.\, the transition structure, for the THI step. 
The molecular structure before and after substitution is compared by aligning the two at the C=C bond. 
When F is introduced at the Y site, the largest change in structure occurs for the transition structure,
while the metastable state is changed most when \ce{CF3} is introduced into site X.
The figure also shows the increase in electron density at the X and Y sites when the fluorine atoms are introduced. The steric hindrance in the transition state for the THI step is increased when fluorine is introduced into site Y, but the main effect of introducing \ce{CF3} at site X is in the metastable state.

The change in the minimum energy path for the THI step, 
calculated using the climbing-image nudged elastic band method 
(CI-NEB),
\cite{millsReversibleWorkTransition1995,henkelmanClimbingImageNudged2000,henkelmanImprovedTangentEstimate2000}
is shown in figure~\ref{fig:data}(i,j). The zero of energy in each case is chosen to be the energy of the state for which the atomic structure changes the least as indicated by the smallest root-mean-square deviation (RMSD). For the fluorine substitution at site Y, this is the transition structure, while for the substitution at site X, it is the metastable state. This illustrates clearly that the main effect of the former is to raise the energy barrier, while the energy of the metastable state with respect to the stable state is not affected much, in agreement with the conclusions reached earlier by Feringa and coworkers~\cite{stackoFluorineSubstitutedMolecularMotors2017}.

However, the \ce{CH3} to \ce{CF3} substitution at site X has a different effect, namely an increase in the energy of the metastable state, while the energy of the transition structure is not changed significantly. As a result, the substitution at site X increases the rotational speed of the molecule, whereas the substitution at site Y slows it down.
The analysis of these effects for \molecule{68} shown in figure~\ref{fig:data}  is representative for the other molecules studied here (see SI).

The increase in the electron density at site Y when the H atom is replaced by an F atom,
is illustrated in figure~\ref{fig:data}(d). 
This leads to greater steric hindrance with the stator during the rotation as is evident from the changes in the transition structure and the height of the energy barrier. 
The distance between the F atom and the H atom in the Z site of the stator is 0.3\,{\AA}\, larger than the distance between the two corresponding H atoms of the base molecule in the transition structure (see SI). 

 
The substitution of CH\textsubscript{3} by CF\textsubscript{3} at site X also increases the electron density at that site, as shown in fig.~\ref{fig:data}(h), but this turns out to introduce an increase in steric hindrance in the metastable state, as can be deduced from the atomic structure changes illustrated in figure~\ref{fig:data}(e-g) and the change in energy along the minimum energy path shown in figure~\ref{fig:data}(j).
The increase in distance between the H atom in the Z site of the stator and the closest F atom at site X as compared to the distance between the two corresponding H atoms in the base molecule is evident mainly in the metastable state, where it is more than 0.5\,{\AA}\,,  while there is little increase in this distance in the transition structure (see figure S1). 
The energy of the transition structure for the backreaction is not affected significantly by the fluorine substitutions (see figure S2).

Further analysis of the steric effect of all the molecules studied here is shown in
figure S3. The average distance between the H atom at site Y of the rotor and atoms of the stator that are within 2.9 Å (the sum of  van der Waals radii of C and H atoms) in either of the two states is shorter in the transition structure than in the metastable state structure, consistent with steric hindrance being larger for the transition state. Conversely, the average distance between the three H atoms of the CH$_3$ group at site X and atoms of the stator that are within 2.9 Å is shorter in the metastable state structure than in the transition structure, so there the steric hindrance is larger for the metastable state. The substitution of H by F at site Y therefore affects the transition state more strongly, while the substitution of CH$_3$ by CF$_3$ at site X has a greater impact on the metastable state.


An important aspect in the design of molecular motors is the difference in wavelength of absorbed light by the stable and metastable states. 
A larger difference is better as it reduces the likelihood of exciting the metastable state and thereby a photoinduced backreaction.
Calculations of spectra for the molecules studied here reveal that this difference increases when the CH\textsubscript{3} to CF\textsubscript{3} substitution at site X is made, 
up to 26 nm for \molecule{60}.
However, it decreases when H is substituted by F at site Y. When both substitutions are made, the difference in absorption wavelength tends to increase (see table S1). 

Since the steric hindrance of the rotor with the Z site of the stator appears to be an important aspect for determining the rate of the THI, another substitution has been studied where the CH at site Z is replaced by an N atom. This has been tested for two of the molecules, \molecule{0} and \molecule{7}. 
In both cases, an increase in the energy of the metastable state by ca. 0.15 eV is found, with little effect on the transition structure. As a result, the half-life of the metastable state is predicted to decrease. However, as only two molecules have been studied, further investigation is required to see how general this effect is. This replacement of CH by N  at site Z is also not found to significantly affect the way the fluorine substitution at site X or Y changes the rate of the THI step.


In summary, 
the rate-limiting step in the rotation of a variety of second-generation molecular motors, the metastable to stable state transition, i.e. THI step, as well as the competing backreaction have been studied theoretically using HTST to evaluate the transition rate based on energy and atomic forces obtained from DFT with the B3LYP hybrid functional. Remarkably good agreement is obtained between the calculated and experimentally measured lifetime of the metastable state establishing the validity of this approach and providing support for predictions made regarding new variants. A CH\textsubscript{3} to CF\textsubscript{3} substitution at the stereogenic center of the rotor is proposed as a way to increase the rotational speed, opposite to the effect of the H to F substitution which has previously been studied by Feringa and co-workers.\cite{stackoFluorineSubstitutedMolecularMotors2017} 
By analysing the minimum energy path of the transitions, calculated using the CI-NEB method,
the effect of these fluorine substitutions can be explained.
The CF\textsubscript{3} substitution turns out to raise the energy of the metastable state, while it has little effect on the transition structure. The H to F substitution, however, raises the energy of the transition structure, while the metastable state is barely affected. 
When both the F and the CF\textsubscript{3} substitutions are made, the rate of the THI step, and thereby the rotational speed, is higher than if only the F substitution is made, but lower than that of the original molecule, thereby providing an opportunity for fine-tuning. 
Furthermore, the calculated wavelength of light absorbed indicates that CF\textsubscript{3} substitution increases the separation between the absorption peaks of the stable state and the metastable state.  This makes it easier to selectively excite the stable state by an appropriate choice of the wavelength of the incident light. The photoisomerization quantum yield is another important factor, but is not addressed here.


\section{Methods}

The rate constants for the thermally activated transitions in the ground electronic state are calculated using the harmonic approximation to transition state theory\cite{wignerTransitionStateMethod1938,vineyardFrequencyFactorsIsotope1957}
\begin{equation}
k_{\rm HTST} = \frac{\prod_i^{3N} \nu_i^{\rm min}}{\prod_i^{3N-1} \nu_i^\ddagger} 
    \exp \left[ -\left(E^\ddagger - E^{\rm min}\right)/k_{\rm B} T \right],
    \label{eqn:htst-rate}
\end{equation}
where $\nu_i^{\rm min}$ and $\nu_i^\ddagger$ refer to vibrational frequency, and $E^{\rm min}$ and $E^\ddagger$ refer to the energy of the initial state minimum and first-order saddle point, respectively. 
A vibrational analysis is, furthermore, used to confirm that the stable and metastable structures are local minima on the energy surface and the transition structures correspond to first-order saddle points. The calculated values of the pre-exponential factor, activation energy and rate constant are given in the SI. 
The half-life of the metastable state shown in figure 3(a) is obtained from the THI and backreaction (thermal inversion, TI) rate constants as $\tau = \ln{2}/(k_{\rm THI}+k_{\rm TI})\,$.
The half-life of the THI step shown in figure 3(b) is $t^{THI}_{1/2}=\ln{2}/k_{THI}$.

The transition structure for the thermally activated steps is determined by calculating the minimum energy path between the metastable-Z state and the stable-Z state.  The configuration of maximal energy along the path corresponds to the saddle point that is used in HTST to represent the transition state. The climbing-image nudged elastic band (CI-NEB) method \cite{millsReversibleWorkTransition1995,henkelmanClimbingImageNudged2000,henkelmanImprovedTangentEstimate2000}
is used
as implemented in the ORCA software\cite{neeseORCAProgramSystem2012,neeseSoftwareUpdateORCA2022} with energy weighted springs\cite{asgeirssonNudgedElasticBand2021}
starting from an initial guess obtained by the sequential image depended pair potential (S-IDPP) method\cite{Schmerwitz2024}.
The location of the images of the molecule along the path are converged to a tolerance of $2.5\cdot 10^{-3}$\,a.u. and $5\cdot 10^{-3}$\,a.u.\ in the root mean square (RMS) and maximum component (MAX) of the force perpendicular to the image tangents, respectively, with one order of magnitude tighter tolerance on the CI. CI-NEB is followed by a first-order saddle point search with  
the CI as the initial guess\cite{asgeirssonNudgedElasticBand2021} and is converged to a tolerance of $5\cdot 10^{-6}$\,a.u., $3\cdot 10^{-4}$\,a.u., $10^{-4}$\,a.u., $4\cdot 10^{-3}$\,a.u., and $2\cdot 10^{-3}$\,a.u.\ for the two-step energy change, the MAX and RMS of the gradient, and the MAX and RMS of the optimization step, respectively. 

The energy of the system and atomic forces are estimated using density functional theory (DFT) calculations with the B3LYP hybrid functional approximation.\cite{leeDevelopmentColleSalvettiCorrelationenergy1988, beckeDensityfunctionalExchangeenergyApproximation1988, beckeDensityFunctionalThermochemistry1993} 
The orbitals are expressed as linear combinations of atomic orbitals using the 6-31G(d,p) basis set\cite{hehreSelfConsistentMolecular2003, weigendAccurateCoulombfittingBasis2006}. 
The electronic structure calculations are converged to thresholds of $10^{-8}$\,a.u., $10^{-7}$\,a.u., $5\cdot 10^{-9}$\,a.u., and $5\cdot 10^{-7}$\,a.u.\ for the two-step energy change, the MAX and RMS of the two-step density change, and the error of the direct inversion in the iterative subspace (DIIS), respectively. The structures of the stable and metastable states of all molecular motors are optimized to a tolerance of $10^{-4}$\,a.u., $3\cdot 10^{-4}$\,a.u., $2\cdot 10^{-3}$\,a.u., and $4\cdot 10^{-3}$\,a.u.\ for the RMS and MAX of the gradient and the RMS and MAX of the optimization step, respectively. The DFT calculations are performed with version 5.0.4 of the ORCA software\cite{neeseORCAProgramSystem2012, neeseSoftwareUpdateORCA2022,Neese2003,Neese2023,Helmich2021}.

While the agreement between calculated and measured values of the lifetime of the metastable state is remarkably good, as illustrated in figure 3(a), our theoretical approach involves several approximations.
The calculations are carried out for isolated molecules while the measurements are carried out on solvated molecules. Also the use of the harmonic approximation to TST rather than full free energy calculations are used, and the energetics are approximated with the B3LYP density functional. The remarkably close agreement between calculated and measured values of the half-life for a wide range of molecular motors using this approach is likely a result of some cancellation of errors.

The optical spectra are obtained with linear-response time-dependent density functional theory\cite{Runge1984, Casida1996} (TDDFT) using the B3LYP functional in the adiabatic approximation. This approach has previously been found to provide results in agreement with experimental measurements for the base MPF molecular rotor\cite{carforaCostEffectiveComputationalStrategy2025}.

Data from the calculations is extracted using the ChemParse software.

\begin{acknowledgement}

This work was funded by the Icelandic Research Fund (grant nos. 239970, 217751 and 2511544)
and the University of Iceland Research Fund. Y.L.A.S.\ acknowledges support by the Max Planck Society.
The calculations were carried out at the IREI computer facility located at the University of Iceland.

\end{acknowledgement}

\begin{suppinfo}
Results on analysis of the changes in atomic distances, minimum energy paths, calculated wavelength of light absorbed, and calculated rates can be found in the Supporting Information (SI).
The data supporting the findings of this work are available for download at Zenodo.
\cite{ZenodoData} 

\end{suppinfo}
\bibliography{main}

\end{document}